\documentclass[journal]{vgtc}                     

\graphicspath{{figures/}{pictures/}{images/}{./}} 

\usepackage{times}                     

\usepackage{tabu}                      
\usepackage{booktabs}                  
\usepackage{lipsum}                    
\usepackage{mwe}                       

\usepackage{mathptmx}                  
\usepackage{amsmath}                   
\usepackage{multirow}
\usepackage{booktabs} 
\usepackage{multirow} 
\usepackage{graphicx} 
\usepackage{colortbl}
\usepackage{float}

\onlineid{0}

\vgtccategory{Research}

\title{DynFOA: Generating First-Order Ambisonics with Conditional Diffusion for Dynamic and Acoustically Complex 360-Degree Videos}

\author{%
  \authororcid{Ziyu Luo\textsuperscript{*}}{0009-0000-2324-0793}, 
  \authororcid{Lin Chen\textsuperscript{*}}{0009-0009-8696-1788}, 
  \authororcid{Qiang Qu}{0000-0002-6648-5050}, 
  \authororcid{Xiaoming Chen\textsuperscript{\textdagger}}{0000-0002-7503-3021}, and 
  \authororcid{Yiran Shen}{0000-0003-1385-1480}
}

\authorfooter{
  \item
    \textsuperscript{*} Ziyu Luo and Lin Chen contributed equally to this work.
  \item
    \textsuperscript{\textdagger} Xiaoming Chen is the corresponding author. 
  \item
  	Ziyu Luo, Lin Chen, and Xiaoming Chen are with School of Computer and Artificial Intelligence, Beijing Technology and Business University, China.
  	E-mail: ziyu.luo@st.btbu.edu.cn; lin.chen@st.btbu.edu.cn; xiaoming.chen@btbu.edu.cn. 
  \item
  	Qiang Qu is with School of Computer Science, The University of Sydney, NSW, Australia. 
  	E-mail: vincent.qu@sydney.edu.au.

  \item Yiran Shen is with School of Software, Shandong University, China.
  	E-mail: yiran.shen@sdu.edu.cn.
}

\abstract{
  Spatial audio is crucial for immersive 360-degree video experiences, yet most 360-degree videos lack it due to the difficulty of capturing spatial audio during recording. Automatically generating spatial audio such as first-order ambisonics (FOA) from video therefore remains an important but challenging problem. In complex scenes, sound perception depends not only on sound source locations but also on scene geometry, materials, and dynamic interactions with the environment. However, existing approaches only rely on visual cues and fail to model dynamic sources and acoustic effects such as occlusion, reflections, and reverberation. 
To address these challenges, we propose DynFOA, a generative framework that synthesizes FOA from 360-degree videos by integrating dynamic scene reconstruction with conditional diffusion modeling. DynFOA analyzes the input video to detect and localize dynamic sound sources, estimate depth and semantics, and reconstruct scene geometry and materials using 3D Gaussian Splatting (3DGS). The reconstructed scene representation provides physically grounded features that capture acoustic interactions between sources, environment, and listener viewpoint. Conditioned on these features, a diffusion model generates spatial audio consistent with the scene dynamics and acoustic context. 
We introduce M2G-360, a dataset of 600 real-world clips divided into MoveSources, Multi-Source, and Geometry subsets for evaluating robustness under diverse conditions. Experiments show that DynFOA consistently outperforms existing methods in spatial accuracy, acoustic fidelity, distribution matching, and perceived immersive experience. 
}


\teaser{
  \centering
  \includegraphics[width=1.0\textwidth]{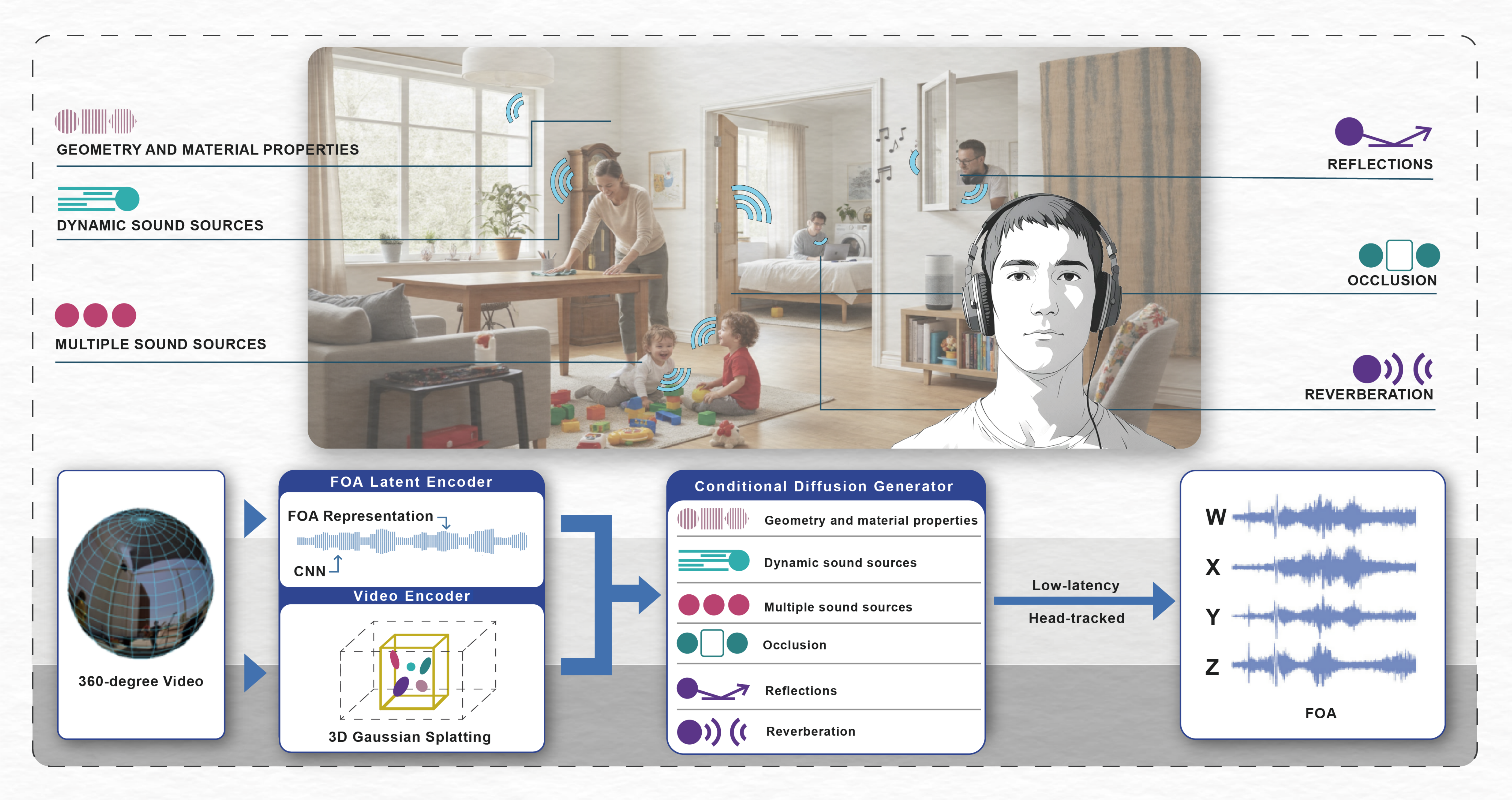}
  \caption{The Pipeline of DynFOA for Immersive Spatial Audio Generation. The high-fidelity spatial audio requires understanding how environmental structures shape sound. \textbf{Top}: In complex scenes, DynFOA captures dynamic multi-source interactions and models their physical propagation effects, including occlusion, reflections, and reverberation. \textbf{Bottom}: By explicitly extracting geometry and material priors via the Video Encoder, the Conditional Diffusion Generator of DynFOA combining with the FOA Latent Encoder can synthesize high-fidelity, head-tracked FOA that deliver a truly immersive and physically coherent auditory experience.
  \label{fig:teaser}
}
}

\graphicspath{{figs/}{figures/}{pictures/}{images/}{./}} 

\usepackage{booktabs}                  
\usepackage{lipsum}                    
\usepackage{mwe}                       
\usepackage{ccicons}                   

\usepackage{mathptmx}                  

\begin{document}


\firstsection{Introduction}

\maketitle

Immersion in Virtual Reality (VR) relies not just on visual fidelity but on the seamless alignment of auditory cues with complex physical environments \cite{begault20003}. While current visual rendering has reached unprecedented levels of realism \cite{kerbl20233d}, generating spatially and physically plausible audio in VR remains a critical bottleneck \cite{chen2020soundspaces}. Unlike traditional audio channels \cite{morgado2018self}, spatial audio, e.g., First-Order Ambisonics (FOA), requires going beyond simple stereo alignment, which demands accurate 3D localization of dynamic sound sources and faithful modeling of complex environmental acoustics, such as occlusion, reflections, and reverberation \cite{NEURIPS2024_32cf311e}, adapting instantly to the listener’s head orientation and human perception.

In immersive 360-degree videos, a major challenge of existing spatial audio rendering methods is their failure to fully exploit the rich geometric and semantic cues. Most FOA-based and hybrid audio-visual models \cite{liuomniaudio,kim2025visage,xie2025sonic4d,zhang2025visaudio} prioritize basic sound source localization, entirely overlooking how the physical environment actively shapes the acoustic field. The state-of-the-art methods, represented by OmniAudio \cite{liuomniaudio} and ViSAGe \cite{kim2025visage}, fail to account for real-world physical material conditions by relying on static and singular visual cues from 360-degree video in advance. These generation approaches often overlook the profound impact of significant spatial environmental factors, such as dynamic objects, room boundaries, and various surface materials on the propagation of sound through obstruction, reflections, and reverberation \cite{NEURIPS2024_32cf311e,chen2020soundspaces}. This omission results in sound fields that lack physical grounding and fail to adapt convincingly to user orientation.

To address these limitations, we propose \textbf{DynFOA}, a FOA generation framework (Figure \ref{fig:teaser}) that leverages the geometric and material information extracted from 360-degree videos. Our method explicitly reconstructs the scene’s 3D geometry from 360-degree video using a pipeline consisting of sound source detection, dense depth estimation, semantic segmentation, and 3D Gaussian Splatting (3DGS) \cite{kerbl20233d} reconstruction. The resulting geometry is augmented with per-surface material properties to derive acoustic features, including occlusion masks, reflection paths, and frequency-dependent reverberation times. These features serve as visual signals for a conditional diffusion-based FOA generator  \cite{kushwaha2025diff}, enabling spatial audio rendering that responds dynamically to both the spatial structure of the scene and the listener’s head orientation. By conditioning the denoising process on real scene geometry, DynFOA produces more physically plausible spatial audio that preserves directionality, distance cues, and environmental characters. To bridge the gap in the FOA generation task based on reconstruction, we further construct the \textbf{M2G-360} dataset, which contains 600 clips involving real-world 360-degree videos. We perform detailed three subsets classification: MoveSources, Multi-Source, and Geometry to verify the model's robustness under various complex environments.

In summary, our main contributions are summarized as follows:

\begin{itemize}
    \item We propose \textbf{DynFOA}, a new conditional diffusion framework for synthesizing high-fidelity FOA from 360-degree videos, fundamentally bridging the gap between spatial visual cues and complex acoustic field generation.
    \item To the best of our knowledge, we are the first to explicitly integrate 3D scene geometry and per-surface material properties into spatial audio synthesis. By leveraging 3DGS, our method rigorously constrains the generative process with physical acoustics, effectively addressing occlusion, reflections, and reverberation in the complex scene.
    \item We establish \textbf{M2G-360}, a meticulously curated novel dataset that addresses the critical void of complex physical constraints in existing benchmarks. It provides specialized subsets to rigorously evaluate spatial audio generation under extreme conditions, which also lays the foundation for the new FOA generation paradigm work in the future.
    \item Extensive experimental evaluations demonstrate that DynFOA outperforms state-of-the-art baselines within spatial accuracy, acoustic fidelity and distribution matching of generated FOA that conforms to human-perceived acoustics.
\end{itemize}

By uniting physics-informed scene reconstruction with conditional diffusion-based generative modeling, DynFOA advances spatial audio rendering beyond purely perceptual alignment toward true audio-visual physical coherence, offering a new pathway for immersive media production in VR and cinematic 360-degree experiences. The demo of DynFOA can be found in the supplementary material.

\section{Related Work}

\subsection{Spatial Audio Rendering}

Spatial audio reproduction in immersive media has long relied on ambisonics to encode 3D sound fields within the spherical harmonic domain \cite{gupta2010three, zotter2019ambisonics}. FOA specifically strikes an optimal balance between spatial resolution and computational efficiency. When coupled with Head-Related Transfer Functions (HRTFs) formatted via standardized protocols \cite{gardner1995hrtf, 969552, majdak2013spatially}, FOA provides the foundational mathematical framework for viewpoint-adaptive rendering in VR, seamlessly balancing perceptual fidelity with real-time processing constraints.

To advance beyond purely signal-driven spatialization, recent paradigms \cite{chen2020soundspaces,chen2022soundspaces} leverage visual and physical priors to guide audio generation. Simulation platforms have highlighted the critical synergy of jointly modeling visual and acoustic cues for comprehensive scene understanding. In the context of 360-degree videos, previous approaches \cite{morgado2018self} successfully extrapolated FOA from monaural audio by utilizing panoramic visual cues. 

Building on this, recent state-of-the-art frameworks have formalized visually-guided spatialization tasks. For instance, these advanced methods \cite{liuomniaudio,zhang2025visaudio} generate immersive spatial audio from omnidirectional visuals, yet they fundamentally assume static, context-free sound sources. Conversely, methods like Sonic4D \cite{xie2025sonic4d} reconstruct dynamic 3D trajectories for viewpoint-adaptive rendering but output strictly binaural audio rather than the more versatile FOA format. While some recent geometry-aware models incorporate depth and structural cues \cite{NEURIPS2024_32cf311e, kim2025visage}, they typically restrict their acoustic modeling to simple source-distance attenuation.

\subsection{Cross-Modal Localization and Datasets }

The cross-modal learning heavily relies on the synergy between visual and auditory modalities, where visual signals inherently constrain sound localization and disambiguate overlapping events \cite{ senocak2018learning}. This alignment is particularly crucial for immersive spatial audio rendering, where complex multi-source environments necessitate robust sound separation. Previous efforts have extensively explored weakly supervised object-sound grounding \cite{mo2022closer} and co-separation frameworks, including speaker-independent audio-visual separation \cite{ephrat2018looking} and lip-synchronized speech extraction \cite{pan2022selective}. 

While large-scale datasets such as AudioSet \cite{gemmeke2017audio}, VGGSound \cite{chen2020vggsound}, and MUSIC \cite{MUSIC} have driven foundational advances in general audio-visual learning, they are fundamentally inadequate for 3D spatial audio synthesis due to their lack of spatial acoustic dimensions. To address this, recent spatialization frameworks have introduced specialized 360-degree video datasets, most notably Sphere360 \cite{liuomniaudio}, and YT-Ambigen \cite{kim2025visage}. However, these spatial audio benchmarks exhibit a critical data void: they overwhelmingly feature simplified, static, or context-agnostic acoustic environments. They critically fail to capture the profound physical interactions with dynamic object occlusion, material-dependent reflections, and multi-source reverberation that characterize real-world physical acoustics.

\subsection{Spatial Audio Generation}

Early neural architectures advanced toward pseudo-binaural or stereo synthesis \cite{xu2021visually} and subsequent models addressed basic mono-to-binaural conversion \cite{parida2022beyond}, they fundamentally lack true 3D volumetric resolution.
Although generation methods, e.g., Points2Sound \cite{lluis2022points2sound} introduced rudimentary geometry and motion cues, these pipelines predominantly oversimplify acoustic propagation.
They routinely assume free-field conditions, modeling only basic distance attenuation while entirely ignoring the complex physical interactions of the surrounding scenarios.

Building upon this, a more recent wave of state-of-the-art generative frameworks \cite{zhang2025visaudio,kim2025visage,kushwaha2025diff,xie2025sonic4d,liuomniaudio} has made notable strides in panoramic and dynamic spatial audio synthesis. However, while these contemporary methods successfully improve the perceptual alignment for generation FOA, they generally formulate spatial audio generation as an unconstrained, data-driven mapping problem. Because they rely primarily on global 2D visual features or simplified distance metrics, they lack the structural priors necessary to model the actual physical mechanics of sound propagation. Specifically, without explicit 3D geometry and surface material analysis, these approaches cannot simulate how sound waves diffract around moving obstacles or how different physical textures absorb and reflect acoustic energy. As a result, they struggle to accurately render dynamic occlusion, material-dependent early reflections, and complex late reverberation. This absence of physical constraints often leads to generated sound fields that lack accurate volumetric depth and struggle to maintain spatial consistency in realistic, reverberant environments.

\subsection{Diffusion Models for Audio Synthesis }

While deep learning-based generative models within foundational diffusion architectures \cite{kongdiffwave} and their spatial extensions \cite{kushwaha2025diff,heydari2025immersediffusion} have established new benchmarks in immersive audio synthesis, they optimize primarily for statistical distribution matching rather than physical accuracy. Operating without explicit 3D structural priors, these frameworks inherently struggle to maintain spatial consistency during complex acoustic interactions. This architectural deficiency persists across diverse generative paradigms: autoregressive models \cite{copet2023simple} offer precise sequence-level control yet cannot simulate multidimensional sound propagation, whereas advanced video-to-audio pipelines \cite{kim2025visage,cheng2025mmaudio} successfully incorporate visual semantics but restrict outputs predominantly to binaural formats. By systematically omitting physics-informed propagation mechanisms, the current generative landscape leaves critical spatial phenomena, such as geometric occlusion and material-dependent reverberation largely unresolved.

\section{Methodology}

\subsection{Problem Definition}

The objective of our work is to enable physics-informed and perceptually coherent immersive perception experiences from 360-degree videos by learning to generate scene-aware spatial audio. Prior approaches \cite{liuomniaudio,kongdiffwave,kim2025visage} to spatial audio rendering often rely on simplified acoustic assumptions, neglecting critical aspects, e.g.,  dynamic sound sources, concurrent source interactions, and propagation effects including occlusion, reflections, and reverberation. Our method directly addresses these challenges by learning from multimodal cues, visual appearance, 3D geometry, and material properties to synthesize FOA that faithfully reflect the physical structure and acoustic conditions of the complex scene. 

During inference, the model inputs the 360-degree video $V$ as the input. All scene-aware acoustic features are derived from $V$ via the designed Video Encoder: 
\begin{equation}
c(V)=\{G(V), M(V), R(V)\}.
\end{equation}
and we learn a mapping:
\begin{equation}
f_\theta: (V, c(V)) \mapsto{S}_{\text{4D}}, 
\end{equation}
where $c(V)$ denotes the scene-aware acoustic 
features. Respectively, $G(V)$ denotes the reconstructed 3D scene geometry, $M(V)$ represents per-surface material properties, and $R(V)$ encodes reverberation and reflections parameters. The learnable function $f_\theta$, parameterized by $\theta$, integrates these modalities to generate a multimodal 4D representation $S_{\text{4D}}$, in which spatially aligned audio and visual cues jointly define the immersive experience.

Solving this problem requires addressing a sequence of coupled sub-tasks across both the visual and audio domains. On the \textbf{visual side}, the generation model must 1) detect and localize sound-emitting and non-emitting objects, 2) estimate depth, 3) perform semantic segmentation, and 4) reconstruct a geometry- and material-aware 3D representation using appropriate techniques \cite{samavati2023deep,zhao2017afully}. On the \textbf{audio side}, the generation model must 1) extract directional cues from FOA channels, 2) encode them into a latent representation, and 3) model complex environments propagation phenomena.

The key challenge lies not only in localizing sound sources but also in handling multiple, dynamic sources within acoustically complex environments~\cite{schissler2016interactive}. By grounding audio generation in geometry and material features, our model captures both static and dynamic elements of the scene. This enables real-time adaptation to source motion, ensuring accurate localization, separation, and a more physically plausible audio field that reflects the spatial relationships inherent in 360-degree visual scenes~\cite{raghuvanshi2010precomputed}.

\begin{figure*}[t]
  \centering
  \includegraphics[width=\textwidth]{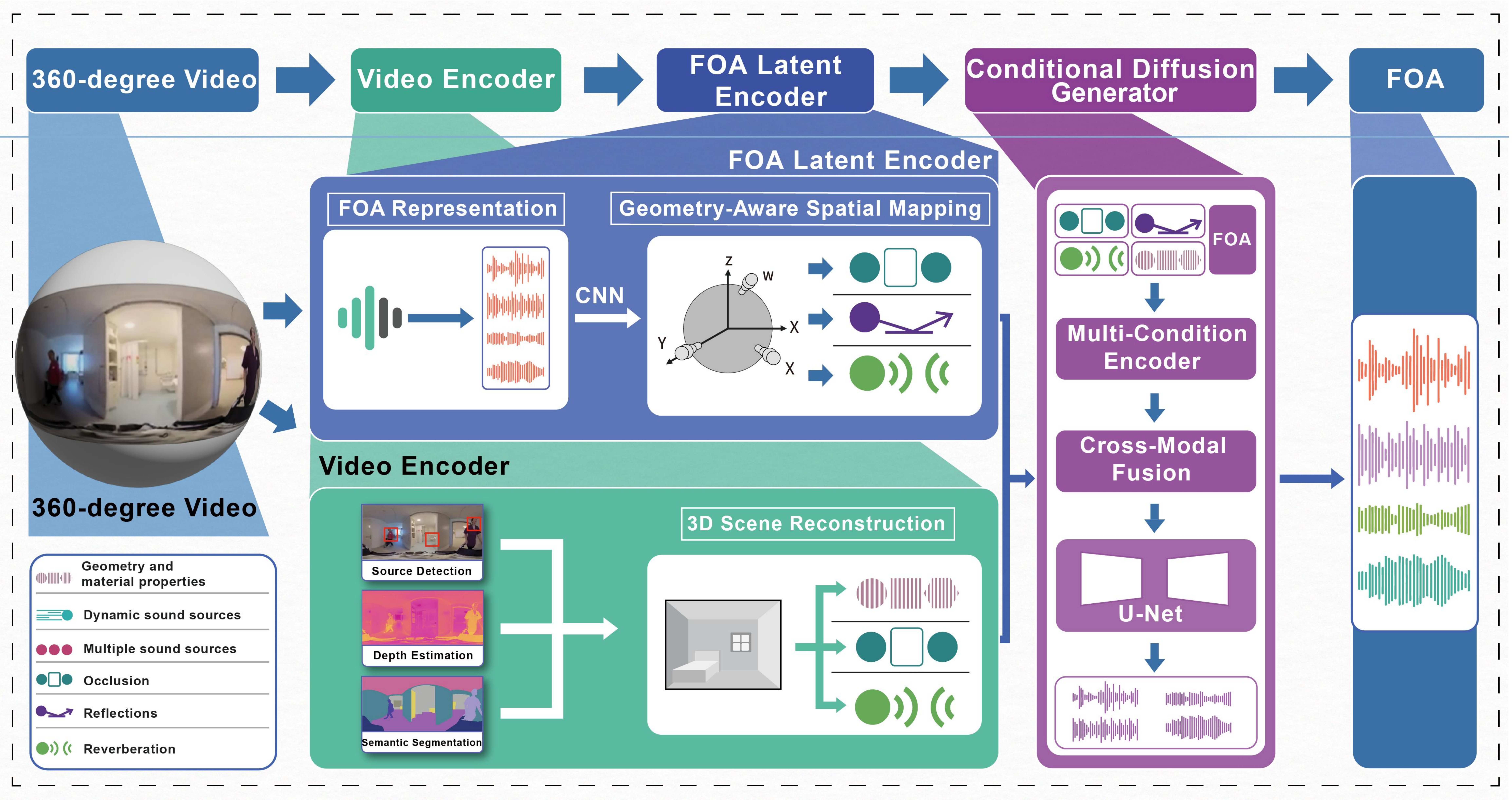}
  \caption{Architecture of the Proposed DynFOA Framework. (1) The \textbf{Video Encoder} reconstructs 3D scene geometry from the 360-degree video via source detection, depth estimation, and semantic segmentation, extracting explicit physical features like occlusion, reflections, and reverberation.
  (2) The \textbf{FOA Latent Encoder} enhances the spatial audio robustness against occlusion, reflections, and reverberation through dynamic sound source processing. Note that this module is utilized only during training to encode ground truth FOA into latent targets. (3) The \textbf{Conditional Diffusion Generator} acts as the core synthesizer. It employs a Multi-Condition Encoder and Cross-Modal Fusion to guide a U-Net denoiser. During the inference, DynFOA drops the FOA Latent Encoder operating purely on video-conditioned diffusion to output high-fidelity spatial audio from the 360-degree video, followed by a pretrained VAE Decoder \cite{kingma2013auto}.}   
  \label{fig:overview}
    \vspace{-1.5em} 
\end{figure*}

\subsection{Framework Overview}

Figure~\ref{fig:overview} illustrates the architecture of the proposed DynFOA framework, which generates dynamic and higher-fidelity FOA from 360-degree videos. The model consists of three main modules, the \textbf{Video Encoder} (see Sec.~\ref{sec:video_encoder}) reconstructs 3D scene geometry and material properties from 360-degree video. It detects and tracks dynamic sound sources, estimates depth, and applies semantic segmentation, producing geometry- and material-aware acoustic features. Furthermore, the \textbf{FOA Latent Encoder} (see Sec.~\ref{sec:audio_encoder}) processes FOA signals into geometry-aware embeddings. Through spectral decomposition, spherical harmonic transformation, and spatial mapping, it captures directional cues, attenuation, and material absorption, while integrating saliency and reverberation features for consistency with the visual scene. Finally, the \textbf{Conditional Diffusion Generator} (see Sec.~\ref{sec:dynfoa_diffusion}) is trained to predict FOA latents encoded from ground-truth FOA during training conditioned on video-derived features. The geometry, material, and propagation cues, along with features of dynamic and multiple sound sources, guide the U-Net denoiser to synthesize FOA that are both physics-informed and perceptually realistic. During inference, the generated FOA is rotated according to the listener’s head orientation and rendered binaurally with HRTFs, enabling low-latency head-tracked playback after FOA generation.

\subsection{Video Encoder}
\label{sec:video_encoder}

This module extracts spatial and semantic information from 360-degree videos to support realistic sound propagation modeling and synchronized spatial audio rendering \cite{tang2021learning}. It operates in three stages including sound source localization and depth estimation, semantic segmentation and scene reconstruction, and features extraction and fusion.

\subsubsection{Sound Source Localization and Depth Estimation}  
We first detect and localize sound-emitting objects in the scene \cite{mo2023audio}. Each source \(i\) is assigned a bounding box \(\hat{b}_i\) and an activity score \(\hat{y}_i\), optimized by:  

\begin{equation}
{L}_{obj} = \sum_i \left( \| b_i - \hat{b}_i \|^2 + (y_i - \hat{y}_i)^2 \right),
\end{equation}

\noindent where \({L}_{obj}\) denotes the joint objective for sound-source localization, \(b_i\) and \(\hat{b}_i\) denote the ground truth and predicted spatial parameters, while \(y_i \in \{0,1\}\) and \(\hat{y}_i \in [0,1]\) represent the true and predicted activity status. This ensures accurate detection and temporal tracking of sound sources.  

The depth estimation then back-projects pixel-level depth $p(u, v)$ into 3D points:  

\begin{equation}
p(u, v) = D(u, v) \left[ \cos(\theta)\cos(\phi), \; \sin(\theta), \; \cos(\theta)\sin(\phi) \right]^T,
\end{equation}

\noindent where \((u,v)\) are image coordinates, \(D(u,v)\) is depth, and \(\theta, \phi\) are the corresponding elevation and azimuth angles. The resulting scene reconstruction serves as the geometric basis for acoustic modeling. Specifically, Truncated Signed Distance Functions (TSDF) \cite{Newcombe2011KinectFusion} 
provide a stable representation of the coarse room layout and large-scale 
structures, while 3DGS \cite{kerbl20233d} preserves fine-grained surface geometry that may be 
lost during volumetric fusion. These two representations therefore 
serve complementary roles: TSDF stabilizes the global scene structure, 
whereas 3DGS acts as a geometric carrier for local surface details used in 
the subsequent extraction of acoustic conditions.   

\subsubsection{Semantic Segmentation and Scene Reconstruction}  

We further apply semantic segmentation to classify scene elements (e.g., walls, floors, furniture). Each class is mapped to frequency-dependent acoustic material properties \cite{van2014influence,ratnarajah2024listen2scene}, enriching the reconstructed geometry with absorption and reflections parameters. These semantic and geometric cues are integrated into a 3D scene model, enabling simulation of occlusion, reflections, and reverberation based on both structure and material characteristics \cite{antani2012interactive}. This ensures that environmental effects such as sound blocking, scattering, and decay are faithfully captured. We assign each semantic class a frequency-band absorption profile using a fixed material lookup table \cite{chen2022soundspaces}, 
and aggregate per-surface parameters onto the reconstructed 3DGS geometry.

\subsubsection{Features Extraction and Fusion}  

To effectively extract spatial and temporal features in the Video Encoder, we introduce optimized CNN \cite{Donahue2017Long} and RNN to address it. These features originated from 360-degree videos, including  3D geometry and material properties, are fused with audio representations to jointly model scene dynamics and acoustic conditions. This fusion allows for the accurate modeling of sound propagation effects, which are crucial for generating physics-informed and perceptually realistic spatial audio. The resulting multimodal features support real-time tracking of dynamic sound sources while incorporating propagation effects from various complex environments. 

\subsection{FOA Latent Encoder}
\label{sec:audio_encoder}

We process the spatial audio signals from FOA channels ($W$, $X$, $Y$, $Z$) via our established FOA Latent Encoder that aligns with the geometry- and material-aware features produced by the Video Encoder. During training, this latent encoding is used as supervision for the conditional diffusion generator. The FOA Latent Encoder outputs a latent code $Z_{\text{FOA}}$ that serves as the training target for the diffusion model. At inference, the sampled latent is decoded into the pretrained VAE Decoder \cite{kingma2013auto} to enhance representations.

\subsubsection{Acoustic Features Extraction}

We begin by extracting the FOA channels, $W$, $X$, $Y$, $Z$, which jointly represent the omnidirectional and directional components of the sound field \cite{liuomniaudio, kushwaha2025diff}. To stabilize training and ensure consistent scaling across channels, z-score normalization is applied by subtracting the mean and dividing by the standard deviation of each channel. This reduces magnitude imbalance and provides a robust foundation for downstream features learning.

From the normalized channels, the optimized CNN \cite{Donahue2017Long} extracts compact representations of spectral and directional patterns. Convolutions over the time frequency domain capture harmonic content and inter-channel correlations, while stacked layers aggregate these into higher-level spatial features. The resulting FOA embeddings form the basis for geometry-aware mapping and materially consistent spatial audio generation.  

\subsubsection{Spatial and Directional Mapping}  
To incorporate structural priors, we modulate FOA features with geometric distance and material-dependent absorption \cite{zotter2019ambisonics}. This accounts for sound attenuation and redirection during propagation:  

\begin{equation}
A_{\text{path}} = \prod_j (1 - \alpha_{mj}) \cdot e^{-\gamma d},
\end{equation}

\noindent where \(A_{\text{path}}\) denotes the overall acoustic amplitude attenuation factor along a propagation path, \(\alpha_{mj}\) is the absorption coefficient of the \(j\) material along the path, \(d\) is the propagation distance, and \(\gamma\) is the air attenuation factor. This formulation supports the modeling of occlusion, reflections, and reverberation.  

Directional information is further captured by projecting FOA features onto a spherical harmonic basis \cite{rafaely2015fundamentals}, yielding a compact representation of spatial energy distributions. This transformation reinforces alignment between audio embeddings and reconstructed scene geometry, enabling accurate reasoning about sound propagation across directions.  

\subsubsection{Acoustic Priors Injection }  

We introduce a cross-modal gating mechanism $a_t$ to modulate acoustic FOA features with visual saliency and text for highlighting perceptually relevant cues:
\begin{equation}
a_t = \sigma(W_{\text{att}} [F_{\text{enc}}; M_{\text{vis}}] + b_{\text{att}}),
\end{equation}

\noindent where \([F_{\text{enc}}; M_{\text{vis}}]\) concatenates encoded FOA features and visual saliency maps, \(W_{\text{att}}\) and \(b_{\text{att}}\) are learnable parameters, and \(\sigma\) is the sigmoid function. This selective amplification refines geometry- and material-aware features, enhancing the simulation of occlusion, reflections, and reverberation.  

Furthermore, we enrich the realistic spatial audio by augmenting FOA features with late reverberation  \cite{ratnarajah2024listen2scene} profiles that capture long-range energy decay and conditional diffusion \cite{kushwaha2025diff}. Estimated from reconstructed geometry and material properties, these profiles complement direct sound and early reflections, yielding acoustically consistent reverberation patterns.  

The FOA Latent Encoder can ultimately produce a geometry and material FOA embedding that is decoded into the FOA channels. This representation preserves directional, spectral, and temporal consistency while remaining aligned with scene geometry, ensuring that the rendered spatial audio is both physics-informed and perceptually coherent.

\subsection{Conditional Diffusion Generator}
\label{sec:dynfoa_diffusion}

Inspired by recent spatial audio researches \cite{kushwaha2025diff,heydari2025immersediffusion}, we employ the Conditional Diffusion Model to synthesize FOA that remains consistent with reconstructed scene geometry and material properties. Operating in the FOA latent domain, the model integrates structural and acoustic cues including occlusion, reflections, and reverberation while accounting for dynamic and multiple sound sources. 


During training, the ground-truth FOA is encoded into a clean latent representation $z_0$. At a randomly sampled diffusion timestep $t$, 
Gaussian noise $\epsilon$ is added to $z_0$ to obtain the noisy latent $z_t$. Conditioned on the video-derived scene and propagation features
$c(V)$, the U-Net denoiser $\epsilon_{\theta}$ is optimized to predict the added noise using the denoising diffusion probabilistic model 
objective: 
\begin{equation}
 {L}_{\text{diff}} = \mathbf{E}_{z_0,\, \epsilon \sim {N}(0,1),\, t}\left[ \left\| \epsilon - \epsilon_{\theta}\left(z_t, t, c(V)\right) \right\|_2^2 \right],
\label{eq:diffusion_loss}
\end{equation}
\noindent where ${L}_{\text{diff}}$ denotes the diffusion training loss, and $c(V)$ aggregates the 
scene geometry, material properties, and acoustic propagation features extracted from the input video. 

\subsubsection{Physics-Informed Conditioning}  

Reconstructed geometry and material attributes provide the foundation for physics-informed synthesis. The reconstruction mesh encodes structural layout and surface orientation, while material properties specify frequency-dependent absorption coefficients \cite{ratnarajah2024listen2scene}. Embedding these features allows DynFOA to account for attenuation, diffraction, and spatial filtering effects, thereby ensuring that the generated FOA is consistent with the reconstructed scene. This condition provides a foundation for handling dynamic and multi-source scenes, enabling accurate simulation between acoustic spatial audio energy and complex environments.

To capture realistic propagation effects, our method is further conditioned on occlusion, reflections, and reverberation \cite{Ho2020Denoising}. Occlusion features are derived from visibility analysis between listener and sources, modulated by material absorption. Occlusion masks are computed by testing visibility between estimated source directions and the listener through the reconstructed geometry 
(e.g., depth-consistent ray casting on the reconstructed scene). Reflections are estimated by tracing geometric paths, providing echo-like cues that enhance spatial depth. Reverberation is represented using frequency-dependent $T_{60}(f)$ \cite{Dal_Santo_2025} curves, which describe late decay characteristics. Together, these cues enrich the conditioning stream, enabling FOA synthesis that incorporates both direct sound and its environmental response.

\subsubsection{Multi-Condition Encoder and Cross-Modal Fusion}
All conditional features are projected into a shared latent space before being injected into the diffusion U-Net. Modulation layers and cross-attention \cite{rombach2022high} mechanisms fuse geometry, material, and propagation cues with intermediate FOA denoising features. This cross-modal fusion guides the denoising trajectory, ensuring that the generated FOA respects physical propagation constraints while maintaining perceptual consistency across time and sources. By explicitly grounding the reverse diffusion process in these 3D physical priors, the network effectively mitigates acoustic hallucinations common in unconstrained generative models. Consequently, the synthesized spatial audio achieves highly accurate localization and realistic reverberation even under severe dynamic occlusion.

\subsubsection{Runtime Rendering and Head-Tracking}  

At inference, the conditional diffusion module generates FOA conditioned on the reconstructed scene and dynamic context. The synthesized FOA signals are rotated according to the listener’s head orientation to maintain spatial alignment under head tracking \cite{zotter2019ambisonics}. Finally, FOA are rendered to binaural signals using HRTFs. Binaural rendering uses a fixed non-individualized HRTFs set from CIPIC \cite{969552}, converted to SOFA format \cite{majdak2013spatially} for standardized access. For each listener pose, the generated FOA is first rotated in the spherical harmonic domain and then rendered to binaural stereo using bilinear interpolation over the discrete HRTFs directions. The same HRTFs set is used for all models and all participants, and no subject-specific HRTFs personalization is applied. We further apply diffuse-field equalization and high-fidelity closed-back headphone compensation for the playback. This runtime process produces immersive spatial audio that adapts to listeners and complex environments including occlusion, reflections and reverberation.

\begin{table*}[t]
\centering
\caption{Benchmarking Comparison Results on the Sphere360. Best results are highlighted in \textbf{Bold}. $\uparrow$/$\downarrow$ indicates that a higher/lower value is better. Note that, because the Ground Truth (GT) is partially absent, we adopt a fitting approach \cite{wang2022open} to predict the GT. This involved approximate supervision using a proxy FOA, extracting the original audio track from 360-degree videos, estimating its spatial cues in the time-frequency domain, and projecting it onto the FOA basis functions to obtain the GT. These proxy signals are injected only for relative comparison
across methods rather than absolute accuracy evaluation. Performance improvement of our DynFOA method across objective and subjective metrics are highlighted with a \colorbox{gray!20}{gray} background against the best baseline OmniAudio.}
\resizebox{\textwidth}{!}{
\begin{tabular}{l | c c c | c c c c | c c}
\toprule
\multirow{2}{*}{Method} & \multicolumn{3}{c|}{Spatial \& Acoustic} & \multicolumn{4}{c|}{Distribution Matching} & \multicolumn{2}{c}{Human Perception} \\
\cmidrule(lr){2-4} \cmidrule(lr){5-8} \cmidrule(lr){9-10}
 & DOA$\downarrow$ & SNR$\uparrow$ & EDT$\downarrow$ & FD$\downarrow$ & KL$\downarrow$ & STFT$\downarrow$ & SI-SDR$\uparrow$ & MOS-SQ$\uparrow$ & MOS-AF$\uparrow$ \\
\midrule
GT (Reference) & - & - & - & - & - & - & - & $4.62 \pm 0.15$ & $4.48 \pm 0.18$ \\
\midrule
ViSAGe \cite{kim2025visage}& 0.48 & 10.95 & 0.16 & 0.36 & 0.70 & 0.58 & 8.35  & $2.62 \pm 0.65$ & $2.45 \pm 0.72$ \\
Diff-SAGe \cite{kushwaha2025diff}& 0.35 & 12.65 & 0.11 & 0.25 & 0.52 & 0.42 & 10.15 & $3.15 \pm 0.52$ & $3.08 \pm 0.58$ \\
MMAudio + SP \cite{cheng2025mmaudio}& 0.26 & 14.80 & 0.08 & 0.19 & 0.40 & 0.28 & 11.75 & $3.65 \pm 0.45$ & $3.22 \pm 0.48$ \\
OmniAudio \cite{liuomniaudio}& 0.19 & 16.85 & 0.06 & 0.14 & 0.31 & 0.21 & 12.68 & $3.96 \pm 0.32$ & $3.82 \pm 0.35$ \\
\midrule
DynFOA (Ours)& \textbf{0.14} & \textbf{18.52} & \textbf{0.04} & \textbf{0.10} & \textbf{0.21} & \textbf{0.14} & \textbf{14.85} & $\mathbf{4.35 \pm 0.22}$ & $\mathbf{4.12 \pm 0.25}$ \\
\midrule
\rowcolor{gray!20}
Improvement& +26.3\% & +9.9\% & +33.3\% & +28.6\% & +32.3\% & +33.3\% & +17.1\% & +9.8\% & +7.9\% \\
\bottomrule
\end{tabular}
}

    \label{sphere360_results}
\end{table*}

\section{Experiment}
\label{sec:exper}

\subsection{Datasets}
\label{datasets}

\subsubsection{Existing Benchmarks}
\label{exist_bench}

In evaluating the generation of FOA, existing research \cite{liuomniaudio,kim2025visage,morgado2020learning} typically relies on several mainstream public benchmark datasets. The most representative is the Sphere360 dataset \cite{liuomniaudio}, which contains a large amount of 360-degree panoramic video and precisely matched spatial audio, and is widely used to evaluate the alignment and reconstruction capabilities of models in multimodal audio-visual spaces. In addition, large-scale datasets, e.g., YT-360 \cite{morgado2020learning}, originating from real streaming platforms YouTube, are often used as important benchmarks for evaluating model generalization capabilities due to their extremely rich coverage of real-world open scenes. While these existing benchmarks have played a crucial role in advancing the field of spatial audio synthesis, they mostly focus on relatively simple acoustic environments or static sound sources, often lacking specific characterization of complex acoustic interactions in physical space, specifically in dynamic occlusion, reflections from diverse materials, and deep reverberation.

\subsubsection{M2G-360 Dataset Construction}
While the existing benchmarks mentioned above provide a solid foundation for general spatial audio evaluation, they lack fine-grained classification and cannot adequately assess complex acoustic phenomena such as dynamic occlusion, material-dependent reflections, and severe reverberation. Inspired by these insights, we establish \textbf{M2G-360}, a newly constructed high-fidelity dataset designed specifically for rigorously evaluating spatial audio reconstruction performance in highly complex acoustic environments. Our dataset draws from a large amount of source material, including the YT-360 \cite{morgado2020learning} and Sphere360 \cite{liuomniaudio}, and is systematically constructed through a hybrid workflow combining rigorous keyword semantic filtering and meticulous human review. Ultimately, this rigorously selected dataset contains 600 high-quality 360-degree video clips. To ensure robust standards for immersive FOA synthesis, each video was carefully normalized: 10 seconds in duration, H.264 encoded, with a minimum resolution of 720p, a frame rate stabilized at 30 FPS, and intrinsically matched to 4-channel FOA audio with a sampling rate of 16kHz and a bit depth of 16bit.

Furthermore, to facilitate highly targeted and multi-dimensional analysis, the M2G-360 clips were systematically divided into three distinct subsets using a content-based filtering approach. Each subset aimed to isolate specific acoustic challenges. Especially, we construct the ``\textbf{MoveSources}'' subset (\textit{128} clips), which isolates highly dynamic scenes containing moving entities such as vehicles and pedestrians, providing controllable conditions for analyzing dynamic occlusion and moving sound propagation. We curate the ``\textbf{Multi-Source}'' subset (\textit{107} clips) to represent complex acoustic scenes with multiple overlapping sound sources, aiming to challenge the limits of the model in resolving simultaneous reflections and complex room reverberation. Finally, the ``\textbf{Geometry}'' subset (\textit{365} clips) captures environments significantly influenced by different structural elements and material properties.

\subsection{Implementation Details}

\subsubsection{Model Training}

To empirically validate the model, all experiments are implemented in PyTorch and deployed on a distributed computing cluster equipped with 8 A100 GPUs (80GB VRAM). Considering the inherently substantial memory footprint of concurrent scene reconstruction and high-fidelity audio synthesis, we construct a robust multi-stage training paradigm with a systematic data partitioning strategy. Specifically, the visual geometric features and material-aware acoustic features, e.g., occlusion masks and reverberation profiles extracted by the Video Encoder, as well as the normalized latent representations of the 4-channel FOA waveforms derived from the FOA Latent Encoder, are pre-computed and cached offline. For all models training, we strictly follow the partitioning strategy from the official Sphere360 \cite{liuomniaudio} protocol. During the conditional diffusion training phase, both the visual and acoustic feature extractors remain strictly frozen. Therefore, DynFOA is not trained end-to-end in the current implementation: the Video Encoder and the FOA Latent Encoder are pre-computed and frozen, while only the Multi-Conditional Encoder and its conditioning projections are optimized during the core training stage. This decoupling strategy effectively alleviates Out-of-Memory bottlenecks, allowing the network to focus exclusively on learning complex Cross-Modal Fusion within the diffusion domain \cite{liu2024audioldm,kongdiffwave,rombach2022high}.

Additionally, the Multi-Conditional U-Net in the Conditional Diffusion Generator is trained for 500,000 steps on this primarily Sphere360-driven partition with an effective global batch size of 128. We optimize the network using the AdamW optimizer \cite{loshchilovdecoupled}, combined with a linear warm-up and cosine annealing learning rate scheduler \cite{loshchilov2017sgdr} to ensure stable convergence across modalities. To improve training throughput while maintaining numerical stability, Automatic Mixed Precision \cite{micikevicius2018mixed} and an Exponential Moving Average of the network weights \cite{songscore} are systematically introduced. By stabilizing the fusion of dynamic visual cues with complex 3D physical priors, our model bridges the gap between 360-degree videos and physically consistent FOA generation.

\begin{table*}[t]
\centering
\caption{Quantitative comparison results on our constructed highly complex M2G-360 with three challenging subsets: MoveSources, Multi-Source, and Geometry to demonstrate the robust ability for the FOA generation task. More specific notes are outlined in Table \ref{sphere360_results}.
}
\resizebox{\textwidth}{!}{
\begin{tabular}{l | c c c | c c c c | c c}
\toprule
\multirow{2}{*}{Method} & \multicolumn{3}{c|}{Spatial \& Acoustic} & \multicolumn{4}{c|}{Distribution Matching} & \multicolumn{2}{c}{Human Perception} \\
\cmidrule(lr){2-4} \cmidrule(lr){5-8} \cmidrule(lr){9-10}
 & DOA$\downarrow$ & SNR$\uparrow$ & EDT$\downarrow$ & FD$\downarrow$ & KL$\downarrow$ & STFT$\downarrow$ & SI-SDR$\uparrow$ & MOS-SQ$\uparrow$ & MOS-AF$\uparrow$ \\
\midrule
\multicolumn{10}{l}{\textit{MoveSources subset}} \\
\midrule
GT (Reference) & - & - & - & - & - & - & - & $4.67 \pm 0.14$ & $4.44 \pm 0.16$ \\
ViSAGe \cite{kim2025visage}& 0.51 & 12.24 & 0.18 & 0.38 & 0.72 & 0.60 & 8.92  & $2.67 \pm 0.68$ & $2.51 \pm 0.73$ \\
Diff-SAGe \cite{kushwaha2025diff}& 0.36 & 12.93 & 0.12 & 0.25 & 0.56 & 0.40 & 10.65 & $3.11 \pm 0.56$ & $3.16 \pm 0.61$ \\
MMAudio + SP \cite{cheng2025mmaudio}& 0.23 & 15.69 & 0.07 & 0.16 & 0.42 & 0.29 & 12.94 & $3.68 \pm 0.42$ & $3.26 \pm 0.46$ \\
OmniAudio \cite{liuomniaudio}& 0.15 & 18.13 & 0.04 & 0.09 & 0.28 & 0.18 & 13.75 & $3.92 \pm 0.29$ & $3.84 \pm 0.31$ \\
\midrule
DynFOA (Ours)& \textbf{0.08} & \textbf{19.92} & \textbf{0.03} & \textbf{0.06} & \textbf{0.17} & \textbf{0.11} & \textbf{15.58} & $\mathbf{4.38 \pm 0.21}$ & $\mathbf{4.17 \pm 0.23}$ \\
\midrule
\rowcolor{gray!20}
Improvement& +46.7\% & +9.9\% & +25.0\% & +33.3\% & +39.3\% & +38.9\% & +13.3\% & +11.7\% & +8.6\% \\
\midrule
\multicolumn{10}{l}{\textit{Multi-Source subset}} \\
\midrule
GT (Reference) & - & - & - & - & - & - & - & $4.58 \pm 0.15$ & $4.49 \pm 0.17$ \\
ViSAGe \cite{kim2025visage}& 0.54 & 11.14 & 0.14 & 0.42 & 0.66 & 0.59 & 7.95  & $2.64 \pm 0.65$ & $2.55 \pm 0.69$ \\
Diff-SAGe \cite{kushwaha2025diff}& 0.39 & 12.16 & 0.10 & 0.28 & 0.50 & 0.41 & 9.86  & $3.11 \pm 0.53$ & $3.18 \pm 0.59$ \\
MMAudio + SP \cite{cheng2025mmaudio}& 0.26 & 14.74 & 0.06 & 0.18 & 0.38 & 0.28 & 11.96 & $3.69 \pm 0.45$ & $3.32 \pm 0.48$ \\
OmniAudio \cite{liuomniaudio}& 0.18 & 16.99 & 0.05 & 0.12 & 0.26 & 0.19 & 12.87 & $4.01 \pm 0.30$ & $3.89 \pm 0.34$ \\
\midrule
DynFOA (Ours)& \textbf{0.12} & \textbf{18.90} & \textbf{0.04} & \textbf{0.08} & \textbf{0.19} & \textbf{0.12} & \textbf{14.47} & $\mathbf{4.34 \pm 0.22}$ & $\mathbf{4.19 \pm 0.24}$ \\
\midrule
\rowcolor{gray!20}
Improvement& +33.3\% & +11.2\% & +20.0\% & +33.3\% & +26.9\% & +36.8\% & +12.4\% & +8.2\% & +7.7\% \\
\midrule
\multicolumn{10}{l}{\textit{Geometry subset}} \\
\midrule
GT (Reference) & - & - & - & - & - & - & - & $4.64 \pm 0.13$ & $4.50 \pm 0.15$ \\
ViSAGe \cite{kim2025visage}& 0.45 & 9.74  & 0.17 & 0.35 & 0.78 & 0.60 & 7.56  & $2.56 \pm 0.66$ & $2.32 \pm 0.75$ \\
Diff-SAGe \cite{kushwaha2025diff}& 0.33 & 12.12 & 0.12 & 0.24 & 0.59 & 0.49 & 9.79  & $3.12 \pm 0.51$ & $2.72 \pm 0.62$ \\
MMAudio + SP \cite{cheng2025mmaudio}& 0.24 & 13.20 & 0.09 & 0.18 & 0.47 & 0.35 & 10.96 & $3.42 \pm 0.44$ & $3.11 \pm 0.49$ \\
OmniAudio \cite{liuomniaudio}& 0.18 & 16.41 & 0.05 & 0.12 & 0.34 & 0.20 & 12.34 & $3.61 \pm 0.33$ & $3.76 \pm 0.36$ \\
\midrule
DynFOA (Ours)& \textbf{0.12} & \textbf{18.37} & \textbf{0.03} & \textbf{0.09} & \textbf{0.27} & \textbf{0.15} & \textbf{15.02} & $\mathbf{4.36 \pm 0.19}$ & $\mathbf{4.03 \pm 0.26}$\\
\midrule
\rowcolor{gray!20}
Improvement& +33.3\% & +11.9\% & +40.0\% & +25.0\% & +20.6\% & +25.0\% & +21.7\% & +20.8\% & +6.4\% \\
\bottomrule
\end{tabular}
}
        \label{ours_results}
        
\end{table*}

\subsubsection{Model Inference}

After the generative network is optimized, the inference and synthesis evaluation phases will be performed on a strictly isolated test dataset. Specifically, this evaluation dataset contains the remainder of the test samples from the Sphere360, as well as all samples (100\%) from our M2G-360 dataset, to better test the model's ability to generate spatial audio under complex acoustic conditions.

Specifically, during inference, only the frozen Video Encoder is executed to extract geometric/material/source features from videos. The FOA Latent Encoder is used only during training to encode GT of FOA into latent targets and is not used at test time. To synthesize FOA audio within a strict latency budget suitable for immersive media, we employ the DPM-Solver++ \cite{lu2025dpm} to bypass the computationally intensive 1000-step reverse process, efficiently generating high-fidelity FOA latent representations in merely 50 denoising steps. These spatial latents are subsequently transformed into continuous 16kHz, 4-channel FOA waveforms using the pre-trained VAE Decoder \cite{liu2023audioldm,liu2024audioldm}. In our current profiling, after the geometry and material features are 
extracted, the remaining DynFOA generation process takes 
110--150\,ms per sample and requires 4.5--6.0\,GB of peak GPU memory. Ultimately, for perceptual evaluation and practical VR applications, the synthesized sound field is dynamically rotated via spherical harmonic matrices to seamlessly align with the listener's head orientation, and then rendered into binaural stereo audio using standard HRTFs.

\subsubsection{Baselines}
To comprehensively evaluate the efficacy and robustness of the proposed \textbf{DynFOA}, we establish a rigorous comparative framework against representative state-of-the-art baselines: (1) \textbf{ViSAGe} \cite{kim2025visage}, a dedicated vision-driven spatial audio generation model. It performs exceptionally well on directional audio synthesis and source localization, providing a robust benchmark for evaluating the directional accuracy and spatial fidelity of our generated FOA. (2) \textbf{Diff-SAGe} \cite{kushwaha2025diff}, a state-of-the-art diffusion-based spatial audio generation model. While it effectively employs a diffusion process for FOA synthesis, it primarily relies on 2D visual cues, serving as an excellent baseline to demonstrate the necessity of explicit 3D geometric modeling. (3) \textbf{MMAudio + Spatialization (SP)} \cite{cheng2025mmaudio}, a combined pipeline that adapts a general-purpose multimodal audio foundation model. We follow the official implementation and add a standard
spherical harmonic spatialization module, and augment it with an audio spatialization component that utilizes spatial angle estimation to upmix the generated audio into the 4-channel FOA format. (4) \textbf{OmniAudio} \cite{liuomniaudio}, a specialized framework for directly converting 360-degree videos into spatial audio. As our primary direct competitor, OmniAudio excels in cross-modal alignment but generates sound fields without physically-grounded scene reconstruction, highlighting the advantages of our scene-based reconstruction method in handling complex occlusions and reverberations. We reproduce the model code from each baseline's official website in turn to ensure the fairness of the comparison. All baselines are retrained on the same training split using their official hyperparameters.

\begin{figure*}[t]
    \centering
    \includegraphics[width=0.98\linewidth]{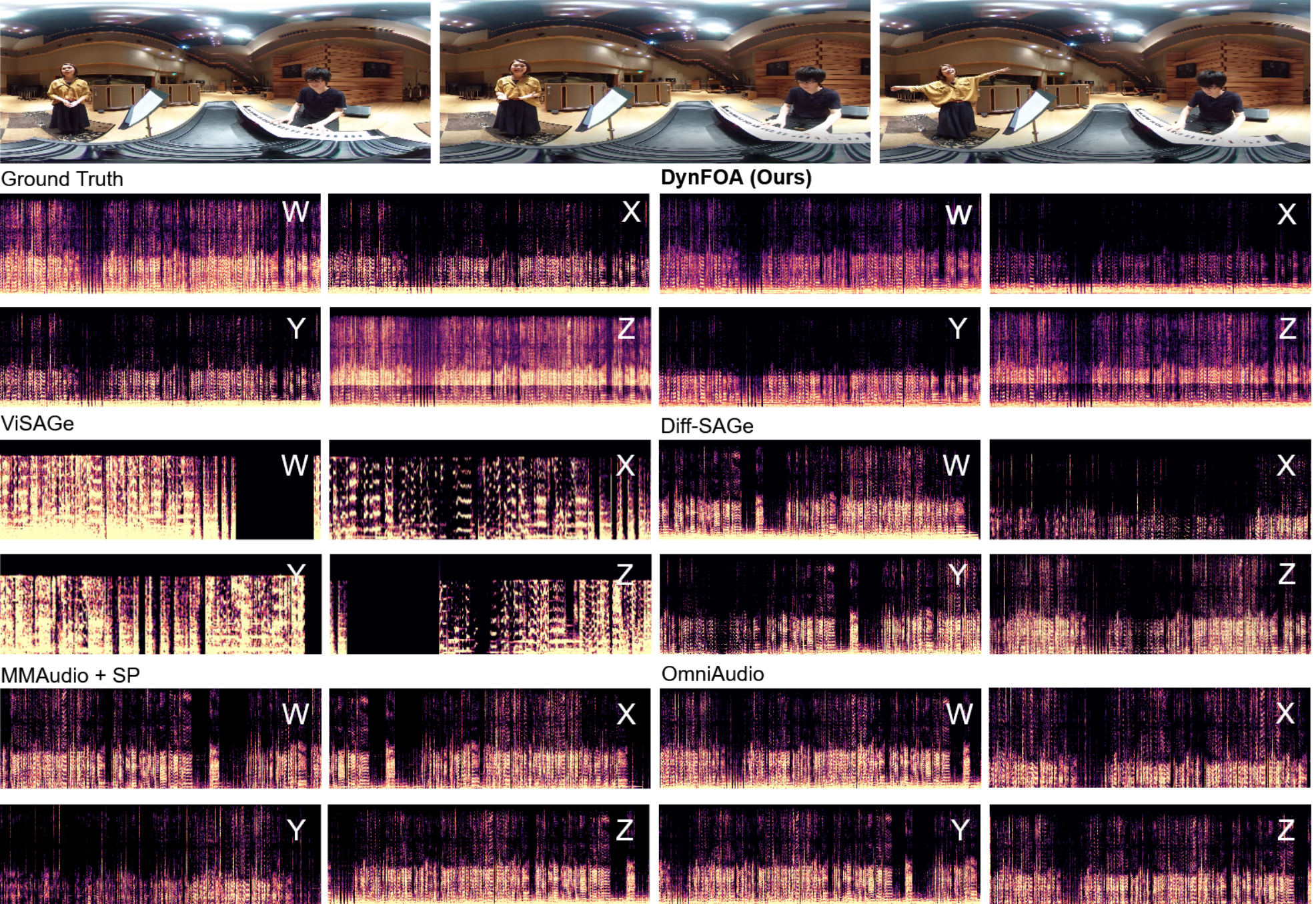}
            \vspace{-1.0em}
    \caption{Visualization comparison of Mel-spectrogram for the FOA channels ($W$, $X$, $Y$, $Z$) in a complex indoor piano environment from M2G-360. Note that ViSAGe is a powerful visually driven FOA baseline, but it primarily relies on visual and directional guidance. In this highly reverberant example, the absence of these explicit conditions may account for its less stable energy distribution across the four channels and its comparatively large deviation from the GT. While state-of-the-art baselines suffer from severe high-frequency attenuation, temporal discontinuities, and loss of inter-channel spatial correlation, our DynFOA successfully reconstructs the full harmonic structure and spatial energy distribution, closely matching the GT. This visually demonstrates that integrating 3D geometry and material priors into the conditional diffusion process effectively prevents acoustic degradation and preserves true physical acoustic coherence in reverberant environments.}
    \label{quan_results}
        \vspace{-1.0em}
\end{figure*}

\subsubsection{Objective Metrics}
We evaluate the FOA generation quality from our DynFOA and baselines along three different dimensions: Spatial Accuracy, Acoustic Fidelity, Distribution Matching. Our experimental objective metrics are specifically designed to validate the effectiveness of various models in handling complex acoustic scenarios, including occlusion, reflections, and reverberation.

\noindent \textbf{Spatial Accuracy} evaluates the directional accuracy of the generated FOA using the Direction of Arrival (DOA) estimation \cite{chen2010introduction}, which measures the angular accuracy of the predicted sound source positions relative to the true source positions in 3D space.

\noindent \textbf{Acoustic Fidelity} assesses different model's ability to capture complex acoustic environments, utilizing the Signal-to-Noise Ratio (SNR) \cite{loizou2007speech} to measure audio clarity in the presence of background noise, and the Early Decay Time (EDT) \cite{cerda2009room} to assess reverberation characteristics in various acoustic spaces, following established practices in spatial audio evaluation. Note, for a fair comparison with non-reconstruction baseline scheme, we extract the $W$ channel from FOA as the standard for EDT calculation \cite{thilakan2025exploring} and fit it with Fourier transform and EDC energy decay curve.

\noindent \textbf{Distribution Matching} considers the similarity of feature distributions between GT and generated FOA under our DynFOA and baselines. According to existing  progress \cite{cheng2025mmaudio,liuomniaudio}, we compute the Fréchet Distance (FD)  features extracted by a pretrained
VGGish-based Spatial Audio Encoder. We further introduce the Short-Time Fourier Transform Error (STFT) \cite{yamamoto2020parallel} to measure spectral reconstruction accuracy, the Scale-Invariant Signal-to-Distortion Ratio (SI-SDR) \cite{jepsen2025study} to assess signal separation quality, and the Kullback-Leibler (KL) \cite{copet2023simple} divergence to evaluate the statistical similarity between generated and reference audio distributions.

\subsubsection{User Study Protocol}
\noindent \textbf{Implementation.} During the human evaluation, participants evaluated a curated subset of 30 video clips (10 seconds each) randomly sampled from the three sub-categories with generated FOA. For each clip, participants experienced the generated spatial audio from DynFOA and the four baselines in a randomized, double-blind order. Upon completion of each clip, the playback paused, and participants used VR controllers to rate the Mean Opinion Score (MOS-SQ) and MOS-AF on a standard 5-point Likert scale (1 = Bad, 5 = Excellent). Here, MOS-SQ measures perceived spatial quality under head-tracked playback, while MOS-AF measures perceived audio-visual alignment between the rendered FOA and videos. They were specifically instructed to pay attention to dynamic source localization and environmental realism (occlusion, reflections, and reverberation). The total duration of the experiment was approximately 40 minutes per participant, including a 5-minute training phase to familiarize them with the questionnaire and rating criteria, and mandatory short breaks to prevent auditory fatigue.

\noindent \textbf{Participants.} We recruited 24 participants (12 males, 12 females) aged between 19 and 35 years (mean = 24.3, SD = 3.1). All participants self-reported normal hearing and normal or corrected-to-normal vision. To ensure a comprehensive evaluation, the cohort included a mix of individuals: 8 had prior professional experience with spatial audio or VR development, while the remaining 16 were novices with only casual or no VR experience.

\noindent \textbf{Equipment.} The subjective evaluations were conducted in a quiet, acoustically treated laboratory environment to eliminate external noise interference. Participants were equipped with a Meta Quest 3 VR headset to view the 360-degree videos with real-time head tracking, paired with high-fidelity closed-back headphone for the binaural rendering of the FOA via standardized HRTFs. The same fixed HRTFs set was used for all participants to avoid introducing subject-specific rendering differences across model comparisons.

\subsection{Main Results}

As detailed in Table \ref{sphere360_results}, DynFOA establishes a new state-of-the-art on Sphere360. While the leading baseline, OmniAudio, relies on static, context-agnostic assumptions, DynFOA explicitly reconstructs 3D geometry and per-surface material properties. This physical grounding enables precise simulation of environmental occlusion and reverberation, driving a 26.3\% reduction in DOA error and a 33.3\% decrease in EDT. Furthermore, compared with unconstrained diffusion model Diff-SAGe that prioritize perceptual fidelity over physical consistency, DynFOA conditions its denoising trajectory on explicit acoustic features. This structural constraint stabilizes the latent distribution, curbing acoustic hallucinations and reducing both STFT error and KL divergence by over 32\%. Ultimately, this rigorous physical conditioning ensures an exceptionally stable, highly immersive user experience, achieving a high score of human perception within MOS-SQ and MOS-AF.

To validate robustness under extreme acoustic conditions, we benchmark DynFOA across three specialized subsets, as reported in Table \ref{ours_results}. For MoveSources subset, integrating video-derived dynamic features with geometry-aware propagation cues enables DynFOA to preserve spatial motion trajectories under severe occlusion, reducing DOA error by a staggering 46.7\%. For Multi-Source subset, the Conditional Diffusion Generator effectively disentangles overlapping sounds via physics-based modeling, improving both STFT error and FD by over 33\%. For the Geometry subset, augmenting spatial information with material properties enables accurate modeling of frequency-dependent reverberation and long-range energy decay, slashing the EDT error by 40.0\%. Finally, the Figure \ref{quan_results} visually corroborates these quantitative gains of selected single sample, demonstrating DynFOA's generated superior FOA performance in a highly reverberant panoramic scene.

\subsection{Ablation Study}

We conduct comprehensive ablation studies on the Sphere360 \cite{liuomniaudio} dataset to verify DynFOA's significant components capability and ensure performance gains are not overfitted to extreme acoustic conditions. Note that DynFOA is trained on this dataset followed by its official protocol, and the model parameters and performance results remain consistent before and after training.

\begin{table}[H]
\centering
\caption{Gradually introducing scene information reduces DOA and
EDT while improving spatial audio frequency-domain stability.}
\label{tab:ablation_scene}
\resizebox{\columnwidth}{!}{
\begin{tabular}{l | c c c c}
\toprule
Variants& DOA$\downarrow$ & SNR$\uparrow$ & EDT$\downarrow$ & FD$\downarrow$ \\
\midrule
Audio-Only                 & 0.26 & 16.50 & 0.09 & 0.22 \\
Audio + Visual& 0.22 & 16.90 & 0.07 & 0.18 \\
Audio + Visual + Depth& 0.18 & 17.50 & 0.05 & 0.14 \\
\midrule
Audio + Visual + Depth + Geo (Ours)& \textbf{0.14} & \textbf{18.52} & \textbf{0.04} & \textbf{0.10} \\
\bottomrule
\end{tabular}
}
\end{table}

\noindent \textbf{Impact of Scene Information.} We compare the audio variant with the gradual addition of our extracted visual priors from 3D scene reconstruction. As shown in Table \ref{tab:ablation_scene}, the transition from the \textbf{Audio-Only} to audio combined with the visual detection (\textbf{+Visual}) results in a notable decrease in both DOA and EDT. This indicates that visual detection benefits reduce localization error and improve temporal stability. Further incorporating the depth prior (\textbf{+Depth}) continues to lower DOA and EDT, showing that depth information effectively enhances spatial consistency of FOA. Finally, combining the geometric scene (\textbf{+Geo}) achieves the lowest DOA and EDT values and the smallest FD, representing the most stable spatial pointing and the least frequency-domain artifacts. These improving results demonstrate that gradually adding scene priors improves spatial accuracy, temporal coherence, and frequency-domain fidelity, bringing the generated audio closer to the high-fidelity.

\begin{table}[H]
\centering
\caption{Backbones of regression diffusion modeling, physical conditioning, and sampling steps for the optimization of generated FOA.}
\label{tab:ablation_diffusion}
\resizebox{\columnwidth}{!}{
\begin{tabular}{l | c c c c}
\toprule
Variants& DOA$\downarrow$ & SNR$\uparrow$ & EDT$\downarrow$ & FD$\downarrow$ \\
\midrule
Regression& 0.28 & 17.10 & 0.08 & 0.25 \\
Simple Diffusion& 0.24 & 16.80 & 0.06 & 0.20 \\
Conditional Diffusion      & 0.17 & 18.00 & 0.05 & 0.13 \\
\midrule
Conditional Diffusion + Steps (Ours)& \textbf{0.14} & \textbf{18.52} & \textbf{0.04} & \textbf{0.10} \\
\bottomrule
\end{tabular}
}
\end{table}

\noindent \textbf{Efficiency of Conditional Diffusion.} We replace our diffusion model backbone to verify the changes in the quality of the generated FOA. As presented in Table \ref{tab:ablation_diffusion}, the \textbf{Regression} module suffers from large DOA and EDT errors as well as higher FD, highlighting the limitations of deterministic regression in maintaining spatial stability. We further introduce the \textbf{Simple Diffusion} backbone changes this behavior: temporal noise modeling reduces DOA and EDT errors, leading to more coherent cross-frame trajectories for generated spatial audio. Finally, we add the \textbf{Conditional Diffusion} with video-derived cues further improves DOA and FD, demonstrating that conditioning primarily enhances spatial coherence and frequency-domain robustness for FOA channels. Increasing the sampling steps (\textbf{+Steps}) achieves the lowest DOA and EDT errors and further reduces FD, indicating that larger step sizes mainly refine acoustic stability and fidelity rather than altering the underlying mechanism. 

\begin{table}[H]
\centering
\caption{Impact of explicitly modeling acoustic propagation effects under complex environments with geometry and materials.}
\label{tab:ablation_propagation}
\resizebox{\columnwidth}{!}{
\begin{tabular}{l | c c c c}
\toprule
Variants& DOA$\downarrow$ & SNR$\uparrow$ & EDT$\downarrow$ & FD$\downarrow$ \\
\midrule
Base Geometry (Free-field)       & 0.20 & 16.50 & 0.12 & 0.22 \\
\quad + Occlusion Masks          & 0.17 & 17.20 & 0.09 & 0.18 \\
\quad + Early Reflections        & 0.15 & 17.80 & 0.07 & 0.14 \\
\midrule
\quad + Late Reverberation (Ours)& \textbf{0.14} & \textbf{18.52} & \textbf{0.04} & \textbf{0.10} \\
\bottomrule
\end{tabular}
}
\end{table}

\noindent \textbf{Impact of sound propagation modeling.} To verify the effectiveness of our core components against complex environments, e.g., occlusion, reflections, and reverberation, we conduct this study on the propagation features of geometry and materials. As shown in Table \ref{tab:ablation_propagation}, a single variant using base 3D geometry without explicitly modeling the material-dependent propagation (\textbf{Free-Field}) yields high EDT and FD errors, as it assumes uniform free-field conditions. Considering the \textbf{Occlusion Masks} based on material absorption immediately improves SNR and slightly refines DOA by correctly attenuating blocked sound paths. Further integrating the \textbf{Early Reflections} significantly enhances spatial depth and spectral fidelity (FD drops to 0.14). Ultimately, incorporating the full frequency-dependent \textbf{Late Reverberation} profiles achieves the optimal configuration. This full model not only delivers the lowest EDT (0.04) by faithfully capturing long-range energy decay but also ensures the highest physical consistency across selective metrics, proving that comprehensive material-aware propagation modeling is indispensable for generated FOA improvement.

\section{Conclusion and Discussion}
\label{sec:conclusion}

In this work, we first propose \textbf{DynFOA}, a generative model that anchors FOA synthesis within the boundaries of real-world acoustic mechanics. Complementing this framework, we establish the \textbf{M2G-360} dataset to additionally evaluate spatial audio generation quality under severe structural and environmental complexities including occlusion, reflections and reverberation. By extracting detailed environmental priors, e.g., surface absorption and spatial depth via the 3DGS, our method successfully bridges the semantic gap between visual layout and wave propagation. Our model can generate high-fidelity spatial audio that conforms to human perception by taking into account geometric and material priors, which traditional methods have failed to address. 

Despite the demonstrated performance, DynFOA has several limitations. First, acoustic properties are estimated through semantic segmentation and 
a predefined material lookup table, which cannot fully capture fine-grained, 
frequency-dependent acoustic characteristics. Second, generalization to 
outdoor and cross-media scenarios remains limited due to incomplete 
geometry, weaker reverberation, complex noise and propagation conditions. Third, the current multi-stage framework is not 
trained end-to-end, which limits joint optimization across scene 
reconstruction, FOA representation learning, and audio generation, while 
reducing memory and computational costs. Future work will explore more 
fine-grained material estimation, broader scenario generalization, and 
memory-efficient end-to-end optimization.  

\section{ACKNOWLEDGMENTS}
This work was supported in part by the Beijing Natural Science Foundation (Grant No. L2611045; Application No. 26L110130, Sub-center Cultivation Project of the Beijing Natural Science Foundation Joint Fund), the National Natural Science Foundation of China (Grant No. 62577004), and the Open Project Program of the State Key Laboratory of Virtual Reality Technology and Systems, Beihang University (Grant No. VRLAB2025C04).

\bibliographystyle{abbrv-doi}

\bibliography{template}

@techreport{begault20003,
  author={Durand R. Begault and Leonard J. Trejo},
  title={{3-D Sound for Virtual Reality and Multimedia}},
  institution={NASA Ames Research Center},
  type={NASA Technical Memorandum},
  number={NASA/TM-2000-209606},
  month= aug,
  year= {2000}
}

@InProceedings{liuomniaudio,
  title = 	 {{{O}mni{A}udio: Generating Spatial Audio from 360-Degree Video}},
  author =       {Liu, Huadai and Luo, Tianyi and Luo, Kaicheng and Jiang, Qikai and Sun, Peiwen and Wang, Jialei and Huang, Rongjie and Chen, Qian and Wang, Wen and Li, Xiangtai and Zhang, Shiliang and Yan, Zhijie and Zhao, Zhou and Xue, Wei},
  booktitle = 	 {Proceedings of the 42nd International Conference on Machine Learning},
  pages = 	 {39060--39084},
  year = 	 {2025},
}

@article{xie2025sonic4d, 
title={{Sonic4D: Spatial Audio Generation for Immersive 4D Scene Exploration}}, 
volume={40}, 
DOI={10.1609/aaai.v40i13.38087},  
number={13}, 
journal={Proceedings of the AAAI Conference on Artificial Intelligence}, 
author={Xie, Siyi and Zhu, Hanxin and Chen, Xinyi and He, Tianyu and Li, Xin and Chen, Zhibo}, 
year={2026}, 
month={Mar.}, 
pages={11087–11095} }

@ARTICLE{gupta2010three,
  author={Gupta, Aastha and Abhayapala, Thushara D.},
  journal={IEEE Transactions on Audio, Speech, and Language Processing}, 
  title={{Three-Dimensional Sound Field Reproduction Using Multiple Circular Loudspeaker Arrays}}, 
  year={2011},
  volume={19},
  number={5},
  pages={1149-1159},
  doi={10.1109/TASL.2010.2082533}}

@article{gardner1995hrtf,
    author = {Gardner, William G. and Martin, Keith D.},
    title = {{HRTF measurements of a KEMAR}},
    journal = {The Journal of the Acoustical Society of America},
    volume = {97},
    number = {6},
    pages = {3907-3908},
    year = {1995},
    month = {06},
    doi = {10.1121/1.412407},
}

@INPROCEEDINGS{969552,
  author={Algazi, V.R. and Duda, R.O. and Thompson, D.M. and Avendano, C.},
  booktitle={Proceedings of the 2001 IEEE Workshop on the Applications of Signal Processing to Audio and Acoustics (Cat. No.01TH8575)}, 
  title={{The CIPIC HRTF database}}, 
  year={2001},
  volume={},
  number={},
  pages={99-102},
  doi={10.1109/ASPAA.2001.969552}}

@inproceedings{majdak2013spatially,
  title={{Spatially oriented format for acoustics: A data exchange format representing head-related transfer functions}},
  author={Majdak, Piotr and Iwaya, Yukio and Carpentier, Thibaut and Nicol, Rozenn and Parmentier, Matthieu and Roginska, Agnieszka and Suzuki, Y{\^o}iti and Watanabe, Kankji and Wierstorf, Hagen and Ziegelwanger, Harald and others},
  booktitle={Audio Engineering Society Convention 134},
  year={2013},
}

@InProceedings{chen2020soundspaces,
author="Chen, Changan
and Jain, Unnat
and Schissler, Carl
and Gari, Sebastia Vicenc Amengual
and Al-Halah, Ziad
and Ithapu, Vamsi Krishna
and Robinson, Philip
and Grauman, Kristen",
title="{{SoundSpaces: Audio-Visual Navigation in 3D Environments}}",
booktitle="Computer Vision -- ECCV 2020",
year="2020",
pages="17--36",
}

@inproceedings{chen2022soundspaces,
author = {Chen, Changan and Schissler, Carl and Garg, Sanchit and Kobernik, Philip and Clegg, Alexander and Calamia, Paul and Batra, Dhruv and Robinson, Philip and Grauman, Kristen},
title = {{SoundSpaces 2.0: a simulation platform for visual-acoustic learning}},
year = {2022},
booktitle = {Proceedings of the 36th International Conference on Neural Information Processing Systems},
}

@inproceedings{morgado2018self,
author = {Morgado, Pedro and Vasconcelos, Nuno and Langlois, Timothy and Wang, Oliver},
title = {{Self-supervised generation of spatial audio for 360° video}},
year = {2018},
booktitle = {Proceedings of the 32nd International Conference on Neural Information Processing Systems},
pages = {360–370},
}

@inproceedings{NEURIPS2024_32cf311e,
author = {Bhosale, Swapnil and Yang, Haosen and Kanojia, Diptesh and Deng, Jiankang and Zhu, Xiatian},
title = {{AV-GS: learning material and geometry aware priors for novel view acoustic synthesis}},
year = {2024},
booktitle = {Proceedings of the 38th International Conference on Neural Information Processing Systems},
}

@INPROCEEDINGS{senocak2018learning,
  author={Senocak, Arda and Oh, Tae-Hyun and Kim, Junsik and Yang, Ming-Hsuan and Kweon, In So},
  booktitle={2018 IEEE/CVF Conference on Computer Vision and Pattern Recognition}, 
  title={{Learning to Localize Sound Source in Visual Scenes}}, 
  year={2018},
  pages={4358-4366},
  doi={10.1109/CVPR.2018.00458}}

@article{ephrat2018looking,
author = {Ephrat, Ariel and Mosseri, Inbar and Lang, Oran and Dekel, Tali and Wilson, Kevin and Hassidim, Avinatan and Freeman, William T. and Rubinstein, Michael},
title = {Looking to listen at the cocktail party: a speaker-independent audio-visual model for speech separation},
year = {2018},
volume = {37},
number = {4},
doi = {10.1145/3197517.3201357},
journal = {ACM Trans. Graph.},
month = jul,
articleno = {112},
numpages = {11},
}

@ARTICLE{pan2022selective,
  author={Pan, Zexu and Tao, Ruijie and Xu, Chenglin and Li, Haizhou},
  journal={IEEE/ACM Transactions on Audio, Speech, and Language Processing}, 
  title={{Selective Listening by Synchronizing Speech With Lips}}, 
  year={2022},
  volume={30},
  number={},
  pages={1650-1664},
  doi={10.1109/TASLP.2022.3153258}}

@INPROCEEDINGS{gemmeke2017audio,
  author={Gemmeke, Jort F. and Ellis, Daniel P. W. and Freedman, Dylan and Jansen, Aren and Lawrence, Wade and Moore, R. Channing and Plakal, Manoj and Ritter, Marvin},
  booktitle={2017 IEEE International Conference on Acoustics, Speech and Signal Processing (ICASSP)}, 
  title={{Audio Set: An ontology and human-labeled dataset for audio events}}, 
  year={2017},
  pages={776-780},
  doi={10.1109/ICASSP.2017.7952261}}

@INPROCEEDINGS{chen2020vggsound,
  author={Chen, Honglie and Xie, Weidi and Vedaldi, Andrea and Zisserman, Andrew},
  booktitle={ICASSP 2020 - 2020 IEEE International Conference on Acoustics, Speech and Signal Processing (ICASSP)}, 
  title={{Vggsound: A Large-Scale Audio-Visual Dataset}}, 
  year={2020},
  pages={721-725},
  doi={10.1109/ICASSP40776.2020.9053174}}

@InProceedings{MUSIC,
author="Zhao, Hang
and Gan, Chuang
and Rouditchenko, Andrew
and Vondrick, Carl
and McDermott, Josh
and Torralba, Antonio",
title="The Sound of Pixels",
booktitle="Computer Vision -- ECCV 2018",
year="2018",
pages="587--604",
}

@article{lluis2022points2sound,
  title={{Points2Sound: from mono to binaural audio using 3D point cloud scenes}},
  author={Llu{\'\i}s, Francesc and Chatziioannou, Vasileios and Hofmann, Alex},
  journal={EURASIP Journal on Audio, Speech, and Music Processing},
  volume={2022},
  number={1},
  pages={33},
  year={2022},
}

@INPROCEEDINGS{parida2022beyond,
  author={Parida, Kranti Kumar and Srivastava, Siddharth and Sharma, Gaurav},
  booktitle={2022 IEEE/CVF Winter Conference on Applications of Computer Vision (WACV)}, 
  title={{Beyond Mono to Binaural: Generating Binaural Audio from Mono Audio with Depth and Cross Modal Attention}}, 
  year={2022},
  pages={2151-2160},
  doi={10.1109/WACV51458.2022.00221}}

@INPROCEEDINGS{xu2021visually,
  author={Xu, Xudong and Zhou, Hang and Liu, Ziwei and Dai, Bo and Wang, Xiaogang and Lin, Dahua},
  booktitle={2021 IEEE/CVF Conference on Computer Vision and Pattern Recognition (CVPR)}, 
  title={{Visually Informed Binaural Audio Generation without Binaural Audios}}, 
  year={2021},
  pages={15480-15489},
  doi={10.1109/CVPR46437.2021.01523}}

@inproceedings{kongdiffwave,
title={{DiffWave: A Versatile Diffusion Model for Audio Synthesis}},
author={Zhifeng Kong and Wei Ping and Jiaji Huang and Kexin Zhao and Bryan Catanzaro},
booktitle={International Conference on Learning Representations},
year={2021},
}

@ARTICLE{liu2024audioldm,
  author={Liu, Haohe and Yuan, Yi and Liu, Xubo and Mei, Xinhao and Kong, Qiuqiang and Tian, Qiao and Wang, Yuping and Wang, Wenwu and Wang, Yuxuan and Plumbley, Mark D.},
  journal={IEEE/ACM Transactions on Audio, Speech, and Language Processing}, 
  title={{AudioLDM 2: Learning Holistic Audio Generation With Self-Supervised Pretraining}}, 
  year={2024},
  volume={32},
  number={},
  pages={2871-2883},
  doi={10.1109/TASLP.2024.3399607}}

@inproceedings{copet2023simple,
author = {Copet, Jade and Kreuk, Felix and Gat, Itai and Remez, Tal and Kant, David and Synnaeve, Gabriel and Adi, Yossi and D\'{e}fossez, Alexandre},
title = {Simple and controllable music generation},
year = {2023},
booktitle = {Proceedings of the 37th International Conference on Neural Information Processing Systems},
}

@InProceedings{cheng2025mmaudio,
    author    = {Cheng, Ho Kei and Ishii, Masato and Hayakawa, Akio and Shibuya, Takashi and Schwing, Alexander and Mitsufuji, Yuki},
    title     = {{MMAudio: Taming Multimodal Joint Training for High-Quality Video-to-Audio Synthesis}},
    booktitle = {Proceedings of the IEEE/CVF Conference on Computer Vision and Pattern Recognition (CVPR)},
    month     = {June},
    year      = {2025},
    pages     = {28901-28911}
}

@article{samavati2023deep,
  title={{Deep learning-based 3D reconstruction: a survey}},
  author={Samavati, Taha and Soryani, Mohsen},
  journal={Artificial Intelligence Review},
  volume={56},
  number={9},
  pages={9175--9219},
  year={2023},
}

@INPROCEEDINGS{zhao2017afully,
  author={Zhao, Cheng and Sun, Li and Stolkin, Rustam},
  booktitle={2017 18th International Conference on Advanced Robotics (ICAR)}, 
  title={{A fully end-to-end deep learning approach for real-time simultaneous 3D reconstruction and material recognition}}, 
  year={2017},
  pages={75-82},
  doi={10.1109/ICAR.2017.8023499}}

@article{schissler2016interactive,
author = {Schissler, Carl and Manocha, Dinesh},
title = {{Interactive Sound Propagation and Rendering for Large Multi-Source Scenes}},
year = {2016},
volume = {36},
number = {1},
doi = {10.1145/2943779},
journal = {ACM Trans. Graph.},
articleno = {114c},
numpages = {12},
month = sep,
}

@inproceedings{raghuvanshi2010precomputed,
author = {Raghuvanshi, Nikunj and Snyder, John and Mehra, Ravish and Lin, Ming and Govindaraju, Naga},
title = {{Precomputed wave simulation for real-time sound propagation of dynamic sources in complex scenes}},
year = {2010},
doi = {10.1145/1833349.1778805},
booktitle = {ACM SIGGRAPH 2010 Papers},
articleno = {68},
numpages = {11},
}

@INPROCEEDINGS{tang2021learning,
  author={Tang, Zhenyu and Meng, Hsien-Yu and Manocha, Dinesh},
  booktitle={2021 IEEE Virtual Reality and 3D User Interfaces (VR)}, 
  title={{Learning Acoustic Scattering Fields for Dynamic Interactive Sound Propagation}}, 
  year={2021},
  pages={835-844},
  doi={10.1109/VR50410.2021.00111}}

@INPROCEEDINGS{mo2023audio,
  author={Mo, Shentong and Tian, Yapeng},
  booktitle={2023 IEEE/CVF Conference on Computer Vision and Pattern Recognition (CVPR)}, 
  title={{Audio-Visual Grouping Network for Sound Localization from Mixtures}}, 
  year={2023},
  pages={10565-10574},
  doi={10.1109/CVPR52729.2023.01018}}

@article{antani2012interactive,
  title={Interactive sound propagation using compact acoustic transfer operators},
  author={Antani, Lakulish and Chandak, Anish and Savioja, Lauri and Manocha, Dinesh},
  journal={ACM Transactions on Graphics (TOG)},
  volume={31},
  number={1},
  pages={1--12},
  year={2012},
}

@article{van2014influence,
title = {{On the influence of frequency-dependent elastic properties in vibro-acoustic modelling of porous materials under structural excitation}},
journal = {Journal of Sound and Vibration},
volume = {333},
number = {24},
pages = {6560-6571},
year = {2014},
doi = {https://doi.org/10.1016/j.jsv.2014.07.032},
author = {C. {Van der Kelen} and P. Göransson and B. Pluymers and W. Desmet},
}

@INPROCEEDINGS{ratnarajah2024listen2scene,
  author={Ratnarajah, Anton and Manocha, Dinesh},
  booktitle={2024 IEEE Conference Virtual Reality and 3D User Interfaces (VR)}, 
  title={{Listen2Scene: Interactive material-aware binaural sound propagation for reconstructed 3D scenes}}, 
  year={2024},
  pages={254-264},
  doi={10.1109/VR58804.2024.00048}}

@INPROCEEDINGS{Newcombe2011KinectFusion,
  author={Newcombe, Richard A. and Izadi, Shahram and Hilliges, Otmar and Molyneaux, David and Kim, David and Davison, Andrew J. and Kohi, Pushmeet and Shotton, Jamie and Hodges, Steve and Fitzgibbon, Andrew},
  booktitle={2011 10th IEEE International Symposium on Mixed and Augmented Reality}, 
  title={{KinectFusion: Real-time dense surface mapping and tracking}}, 
  year={2011},
  pages={127-136},
  doi={10.1109/ISMAR.2011.6092378}}

@article{kerbl20233d,
author = {Kerbl, Bernhard and Kopanas, Georgios and Leimkuehler, Thomas and Drettakis, George},
title = {{3D Gaussian Splatting for Real-Time Radiance Field Rendering}},
year = {2023},
volume = {42},
number = {4},
doi = {10.1145/3592433},
journal = {ACM Trans. Graph.},
month = jul,
articleno = {139},
numpages = {14},
}

@article{Donahue2017Long,
author = {Donahue, Jeff and Hendricks, Lisa Anne and Rohrbach, Marcus and Venugopalan, Subhashini and Guadarrama, Sergio and Saenko, Kate and Darrell, Trevor},
title = {{Long-Term Recurrent Convolutional Networks for Visual Recognition and Description}},
year = {2017},
volume = {39},
number = {4},
doi = {10.1109/TPAMI.2016.2599174},
journal = {IEEE Trans. Pattern Anal. Mach. Intell.},
month = apr,
pages = {677–691},
}

@book{zotter2019ambisonics,
  title={{Ambisonics: A practical 3D audio theory for recording, studio production, sound reinforcement, and virtual reality}},
  author={Zotter, Franz and Frank, Matthias},
  year={2019},
  publisher={Springer}
}

@inproceedings{morgado2020learning,
author = {Morgado, Pedro and Li, Yi and Vasconcelos, Nuno},
title = {{Learning representations from audio-visual spatial alignment}},
year = {2020},
booktitle = {Proceedings of the 34th International Conference on Neural Information Processing Systems},
articleno = {397},
numpages = {12},
}

@book{rafaely2015fundamentals,
  title={{Fundamentals of spherical array processing}},
  author={Rafaely, Boaz},
  volume={8},
  year={2015},
  publisher={Springer}
}

@INPROCEEDINGS{kushwaha2025diff,
  author={Kushwaha, Saksham Singh and Ma, Jianbo and Thomas, Mark R. P. and Tian, Yapeng and Bruni, Avery},
  booktitle={ICASSP 2025 - 2025 IEEE International Conference on Acoustics, Speech and Signal Processing (ICASSP)}, 
  title={{Diff-SAGe: End-to-End Spatial Audio Generation Using Diffusion Models}}, 
  year={2025},
  pages={1-5},
  doi={10.1109/ICASSP49660.2025.10888882}}

@INPROCEEDINGS{heydari2025immersediffusion,
  author={Heydari, Mojtaba and Souden, Mehrez and Conejo, Bruno and Atkins, Joshua},
  booktitle={ICASSP 2025 - 2025 IEEE International Conference on Acoustics, Speech and Signal Processing (ICASSP)}, 
  title={{ImmerseDiffusion: A Generative Spatial Audio Latent Diffusion Model}}, 
  year={2025},
  pages={1-5},
  doi={10.1109/ICASSP49660.2025.10889311}
  }

@INPROCEEDINGS{rombach2022high,
  author={Rombach, Robin and Blattmann, Andreas and Lorenz, Dominik and Esser, Patrick and Ommer, Björn},
  booktitle={2022 IEEE/CVF Conference on Computer Vision and Pattern Recognition (CVPR)}, 
  title={{High-Resolution Image Synthesis with Latent Diffusion Models}}, 
  year={2022},
  pages={10674-10685},
  doi={10.1109/CVPR52688.2022.01042}
  }

@BOOK{chen2010introduction,
  author={Chen, Zhizang and Gokeda, Gopal and Yu, Yiqiang},
  title={{Introduction to Direction-of-Arrival Estimation}},
  year={2010},
  pages={},
  publisher={Artech},
 }

@book{loizou2007speech,
author = {Loizou, Philipos C.},
title = {Speech Enhancement: Theory and Practice},
year = {2013},
publisher = {CRC Press, Inc.},
}

@article{cerda2009room,
title = {Room acoustical parameters: A factor analysis approach},
journal = {Applied Acoustics},
volume = {70},
number = {1},
pages = {97-109},
year = {2009},
doi = {https://doi.org/10.1016/j.apacoust.2008.01.001},
author = {S. Cerdá and A. Giménez and J. Romero and R. Cibrián and J.L. Miralles},
}

@INPROCEEDINGS{yamamoto2020parallel,
  author={Yamamoto, Ryuichi and Song, Eunwoo and Kim, Jae-Min},
  booktitle={ICASSP 2020 - 2020 IEEE International Conference on Acoustics, Speech and Signal Processing (ICASSP)}, 
  title={{Parallel Wavegan: A Fast Waveform Generation Model Based on Generative Adversarial Networks with Multi-Resolution Spectrogram}}, 
  year={2020},
  pages={6199-6203},
  doi={10.1109/ICASSP40776.2020.9053795}}

@INPROCEEDINGS{jepsen2025study,
  author={Jepsen, Simon Dahl and Christensen, Mads Græsbøll and Jensen, Jesper Rindom},
  booktitle={2025 IEEE Automatic Speech Recognition and Understanding Workshop (ASRU)}, 
  title={{A Study of the Scale Invariant Signal to Distortion Ratio in Speech Separation with Noisy References*}}, 
  year={2025},
  pages={1-8},
  doi={10.1109/ASRU65441.2025.11434756}}

@inproceedings{kim2025visage,
title={{Vi{SAG}e: Video-to-Spatial Audio Generation}},
author={Jaeyeon Kim and Heeseung Yun and Gunhee Kim},
booktitle={The Thirteenth International Conference on Learning Representations},
year={2025},
}

@InProceedings{liu2023audioldm,
  title = 	 {{{A}udio{LDM}: Text-to-Audio Generation with Latent Diffusion Models}},
  author =       {Liu, Haohe and Chen, Zehua and Yuan, Yi and Mei, Xinhao and Liu, Xubo and Mandic, Danilo and Wang, Wenwu and Plumbley, Mark D},
  booktitle = 	 {Proceedings of the 40th International Conference on Machine Learning},
  pages = 	 {21450--21474},
  year = 	 {2023},
  volume = 	 {202},
  month = 	 {23--29 Jul},
}

@INPROCEEDINGS{wang2022open,
  author={Wang, Weiyao and Feiszli, Matt and Wang, Heng and Malik, Jitendra and Tran, Du},
  booktitle={2022 IEEE/CVF Conference on Computer Vision and Pattern Recognition (CVPR)}, 
  title={{Open-World Instance Segmentation: Exploiting Pseudo Ground Truth From Learned Pairwise Affinity}}, 
  year={2022},
  pages={4412-4422},
  doi={10.1109/CVPR52688.2022.00438}}

@article{thilakan2025exploring,
    author = {Thilakan, Jithin and B T, Balamurali and Colella Gomes, Otavio and Chen, Jer-Ming and Kob, Malte},
    title = {{Exploring the role of room acoustic environments in the perception of musical blending}},
    journal = {The Journal of the Acoustical Society of America},
    volume = {157},
    number = {2},
    pages = {738-754},
    year = {2025},
    month = {02},
    doi = {10.1121/10.0035563},
}

@inproceedings{loshchilovdecoupled,
title={{Decoupled Weight Decay Regularization}},
author={Ilya Loshchilov and Frank Hutter},
booktitle={International Conference on Learning Representations},
year={2019},
}

@inproceedings{loshchilov2017sgdr,
title={{{SGDR}: Stochastic Gradient Descent with Warm Restarts}},
author={Ilya Loshchilov and Frank Hutter},
booktitle={International Conference on Learning Representations},
year={2017},
}

@inproceedings{micikevicius2018mixed,
title={{Mixed Precision Training}},
author={Paulius Micikevicius and Sharan Narang and Jonah Alben and Gregory Diamos and Erich Elsen and David Garcia and Boris Ginsburg and Michael Houston and Oleksii Kuchaiev and Ganesh Venkatesh and Hao Wu},
booktitle={International Conference on Learning Representations},
year={2018},
}

@inproceedings{songscore,
title={{Score-Based Generative Modeling through Stochastic Differential Equations}},
author={Yang Song and Jascha Sohl-Dickstein and Diederik P Kingma and Abhishek Kumar and Stefano Ermon and Ben Poole},
booktitle={International Conference on Learning Representations},
year={2021},
}

@article{lu2025dpm,
  title={{Dpm-solver++: Fast solver for guided sampling of diffusion probabilistic models}},
  author={Lu, Cheng and Zhou, Yuhao and Bao, Fan and Chen, Jianfei and Li, Chongxuan and Zhu, Jun},
  journal={Machine Intelligence Research},
  volume={22},
  number={4},
  pages={730--751},
  year={2025},
}

@article{zhang2025visaudio,
  title={{ViSAudio: End-to-End Video-Driven Binaural Spatial Audio Generation}},
  author={Zhang, Mengchen and Chen, Qi and Wu, Tong and Liu, Zihan and Lin, Dahua},
  journal={arXiv preprint arXiv:2512.03036},
  year={2025}
}

@inproceedings{mo2022closer,
author = {Mo, Shentong and Morgado, Pedro},
title = {{A closer look at weakly-supervised audio-visual source localization}},
year = {2022},
booktitle = {Proceedings of the 36th International Conference on Neural Information Processing Systems},
articleno = {2720},
numpages = {13},
}

@article{kingma2013auto,
  title={{Auto-encoding variational bayes}},
  author={Kingma, Diederik P and Welling, Max},
  journal={arXiv preprint arXiv:1312.6114},
  year={2013}
}

@article{Dal_Santo_2025,
   title={Optimizing tiny colorless feedback delay networks},
   volume={2025},
   DOI={10.1186/s13636-025-00401-w},
   number={1},
   journal={EURASIP Journal on Audio, Speech, and Music Processing},
   author={Dal Santo, Gloria and Prawda, Karolina and Schlecht, Sebastian J. and Välimäki, Vesa},
   year={2025},
   month=Mar 
}

@inproceedings{Ho2020Denoising,
author = {Ho, Jonathan and Jain, Ajay and Abbeel, Pieter},
title = {Denoising diffusion probabilistic models},
year = {2020},
booktitle = {Proceedings of the 34th International Conference on Neural Information Processing Systems},
articleno = {574},
numpages = {12},
}

\end{document}